\documentclass[a4paper,11pt]{article}
\pdfoutput=1 

\usepackage{jinstpub} 

\title{Characterisation and performance of the PADME electromagnetic calorimeter}

\collaboration{PADME collaboration}

\author[a]{P. Albicocco,}
\author[b]{J. Alexander,}
\author[a]{F. Bossi,}
\author[c]{P. Branchini,}
\author[a]{B. Buonomo,}
\author[a]{C. Capoccia,}
\author[a]{E. Capitolo,}
\author[d]{G. Chiodini,}
\author[d,e]{A.P. Caricato,}
\author[a]{R. de Sangro,}
\author[a]{C. Di Giulio,}
\author[a]{D. Domenici,}
\author[f]{F. Ferrarotto,}
\author[a]{G. Finocchiaro,}
\author[f,g]{S. Fiore,} 
\author[a]{L.G. Foggetta,}
\author[b]{A. Frankenthal,}
\author[h,a]{G. Georgiev,}
\author[a]{A. Ghigo,}
\author[a]{F. Giacchino,}
\author[a]{P. Gianotti,}
\author[h]{S. Ivanov,}
\author[h,a]{V. Kozhuharov,}
\author[f]{E. Leonardi,}
\author[i]{B. Liberti,}
\author[j,f]{E. Long,}
\author[d,e]{M. Martino,}
\author[d,e]{I. Oceano,}
\author[d,e]{F. Oliva,}
\author[j,f]{G.C. Organtini,}
\author[f,j,1]{G. Piperno,\note{Corresponding author.}}
\author[j,f]{M. Raggi,} 
\author[f]{F. Safai Tehrani,}
\author[a]{I. Sarra,}
\author[a]{B. Sciascia,}
\author[h]{R. Simeonov,}
\author[a]{A. Saputi,}
\author[a]{T. Spadaro,}
\author[d,e]{S. Spagnolo,}
\author[a]{E. Spiriti,}
\author[c]{D. Tagnani,}
\author[a,k]{C. Taruggi,}
\author[h]{L. Tsankov,}
\author[f]{P. Valente,}
\author[a]{and E. Vilucchi}

\affiliation[a]{INFN Laboratori Nazionali di Frascati,\\Via E. Fermi 54, 00044 Frascati, Italy}
\affiliation[b]{Department of Physics, Cornell University,\\109 Clark Hall, Ithaca, NY 14853, U.S.A.}
\affiliation[c]{INFN Sezione di Roma 3, Via della Vasca Navale, 84, 00146 Roma RM, Italy}
\affiliation[d]{INFN Sezione di Lecce,\\Via Provinciale per Arnesano, 73100 Lecce, Italy}
\affiliation[e]{Dipartimento di Matematica e Fisica, Universit\`a del Salento,\\Via Provinciale per Arnesano, 73100 Lecce, Italy}
\affiliation[f]{INFN Sezione di Roma,\\p.le Aldo Moro 5, 00185 Roma, Italy}
\affiliation[g]{ENEA centro ricerche Frascati,\\Via Enrico Fermi 45, 00044 Frascati (Roma), Italy}
\affiliation[h]{University of Sofia St. Kl. Ohridski,\\15 Tsar Osvoboditel Blvd, 1504 Sofia, Bulgaria}
\affiliation[i]{INFN Sezione di Roma Tor Vergata,\\Via della Ricerca Scientifica 1, 00133 Roma, Italy}
\affiliation[j]{Dipartimento di Fisica, Sapienza Universit\`a di Roma,\\P.le Aldo Moro 5, 00185 Roma, Italy}
\affiliation[k]{Dipartimento di Fisica Univerisit\`a di  Roma Tor Vergata,\\Via della Ricerca Scientifica 1, 00133 Roma, Italy}

\emailAdd{gabriele.piperno@roma1.infn.it}

\abstract{The PADME experiment at the LNF Beam Test Facility searches
for dark photons produced in the annihilation of positrons with the electrons of a fixed target. 
The strategy is to look for the reaction $e^{+}+e^{-}\rightarrow \gamma+A'$,
where $A'$ is the dark photon, which cannot be observed directly
or via its decay products. The electromagnetic calorimeter plays a key role in the experiment by measuring the energy and position of the final-state $\gamma$. The missing four-momentum carried away by the $A'$ can be evaluated from this information and the particle mass inferred. This paper presents the design, construction, and calibration of the PADME's electromagnetic calorimeter. The results achieved in terms of equalisation,
detection efficiency and energy resolution during the first phase
of the experiment demonstrate the effectiveness of the various tools
used to improve the calorimeter performance with respect to earlier prototypes.}


\begin{document}
\maketitle
\flushbottom

\section{Introduction}

A possible explanation for the elusiveness of dark matter (DM) is that
it interacts with standard model (SM) particles only by means of a
mediator that feebly couples to them. One of the simplest theoretical models based on this assumption introduces a new $U(1)$ gauge symmetry with a vector boson mediator, the dark photon (DP), or $A'$ \cite{DP 1,DP 2}. 
SM particles are neutral under this new symmetry but the $A'$ could faintly interact with them because of a kinetic mixing with the ordinary photon.
The strength of this interaction is given by an effective charge $\varepsilon q$, where $q$ is the electric charge of the particle and $\varepsilon$ is the kinetic mixing coefficient. 
\par
The Positron Annihilation into Dark Matter Experiment (PADME), ongoing at the Laboratori Nazionali di Frascati (LNF) of the Istituto Nazionale di Fisica Nucleare (INFN), aims to explore this theoretical scenario. 
For this purpose, a dedicated detector was built, whose central component is an electromagnetic calorimeter made of crystals of bismuth germanium oxide (BGO). The calorimeter has high granularity and excellent detection efficiency for photons with energy between $100$ and $500\,\text{MeV}$. Sensitivity to the coupling constant down to $1\cdot 10^{-3}$ could be reached for values of $m_{A'} \le23.7\,\text{MeV/c}^2$ collecting $10^{13}$ positron on target.
In the following sections, a short description of the experimental technique is given (section \ref{exp}), followed by a detailed description of the design solutions adopted for the calorimeter (section \ref{calo}), and finally a report of the performance reached in terms of efficiency and energy resolution in the 2018-19 PADME commissioning run (section \ref{calib}).

\section{The PADME experiment}
\label{exp}
PADME is hosted at the Beam Test Facility of LNF \cite{BTF1,BTF2}. 
It looks for a DP produced via the reaction $e^{+}+e^{-}\rightarrow \gamma+A'$ using
positrons accelerated to $550\,\text{MeV}$ by the laboratory's
LINAC, impinging on the electrons of a $100\,\mu\text{m}$ thick active diamond target, 
able to measure beam position and intensity \cite{Target}. 

The experimental technique relies
on the measurement of the missing mass of the final states with a single photon. 
By measuring its energy and the direction of flight with a granular electromagnetic calorimeter (ECal), it is possible
to measure $m_{A'}^{2}$ as the square of the missing mass.

\begin{figure}
\begin{centering}
\includegraphics[width=1.0\textwidth]{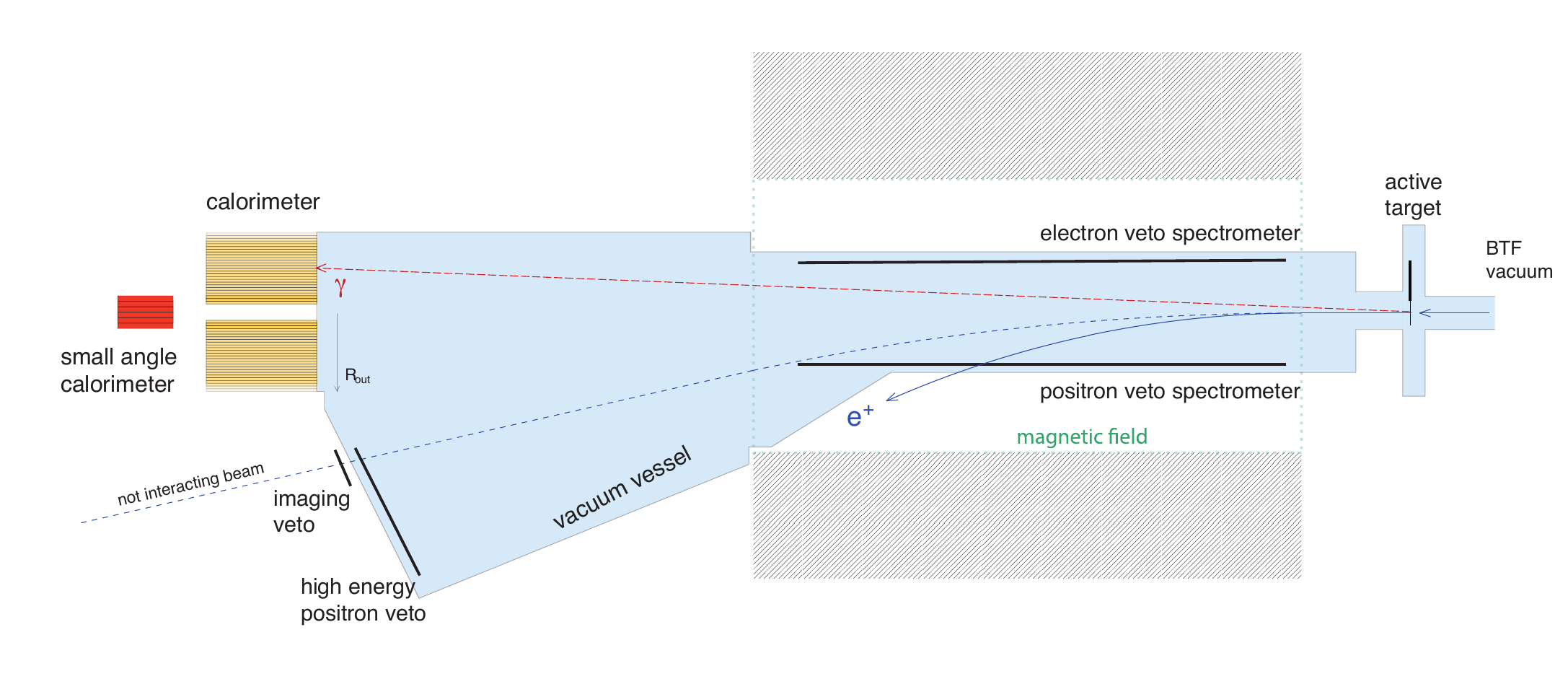}
\par\end{centering}
\caption{\label{fig:PADME_detector} Top view of the PADME detector (not to scale). From right to left: active target, dipole magnet (with $e^{+}/e^{-}$ vetoes inside), high-energy positron veto and the calorimeters.}
\end{figure}

In addition to the ECal, the setup consists of a dipole magnet to deflect the beam away from the calorimeter,
a charged particle veto system (made of a positron veto, electron
veto and high-energy positron veto \cite{Veto1,Veto2}) 
to identify positrons that lose part of their energy through Bremsstrahlung, and a fast small
angle calorimeter (SAC) \cite{SAC} to detect and veto Bremsstrahlung photons in the forward direction. 
A schematic drawing of the detector is shown in figure~\ref{fig:PADME_detector}.

A more detailed description of the PADME detector can be found in \cite{PADME}.

\section{The PADME electromagnetic calorimeter}
\label{calo}

The segmented PADME electromagnetic calorimeter was built by reshaping BGO crystals recovered from
the endcaps of the calorimeter of the L3 experiment \cite{L3}. 
The following sub-sections present a general description of the calorimeter, the production procedure of the scintillating units (SUs) (i.e. the assembly of a crystal and its photomultiplier tube (PMT)), the solution adopted for the signal digitisation, and finally the trigger systems used to collect data.

\subsection{Detector characteristics}

ECal consists of $616$ BGO crystals ($2.1\times2.1\times23.0\,\text{cm}^{3}$) arranged in a cylindrical shape ($\approx29\,\text{cm}$ external radius) and features a central square hole ($5\times5$ crystals) behind which the SAC is placed. 

The size of the calorimeter is a compromise between large acceptance, good angular resolution, and the space available in the experimental hall. 
The distance between the target and the ECal is $3.46\,\text{m}$, determining  an angular coverage of $[15.7,82.1]\,\text{mrad}$.
For the SAC, the covered region is $[0,18.9]\,\text{mrad}$.
The numbers in brackets report respectively half of the minimum and of the maximum opening angle of a cone with vertex at the target and base contained by the front face of the corresponding calorimeter.

The main requirements for the ECal crystals are a small Moli\`ere radius, and an energy resolution of roughly $2\%/\sqrt{E}$. BGO was chosen because it satisfies these conditions and because BGO crystals were available for reuse from the L3 experiment~\cite{L3}.

The crystal transverse dimensions were chosen to provide excellent position resolutions, while the longitudinal dimensions of the crystals, slightly more than $20$ radiation lengths, were imposed by the original L3 crystal length. 

The central square hole allows the passage of forward emitted Bremsstrahlung photons, detected by the SAC.
The high rate of these photons would flood the inner ECal crystals. On the contrary,  being based on Cherenkov radiator (PbF$_2$), the SAC is able to sustain particle rates up to hundreds of MHz,
thanks to a signal length of only $3\,\text{ns}$, much shorter with respect to BGO scintillation light decay time of $\sim300\,\text{ns}$ \cite{PDG}.   

\begin{figure}
\begin{centering}
\includegraphics[height=5cm]{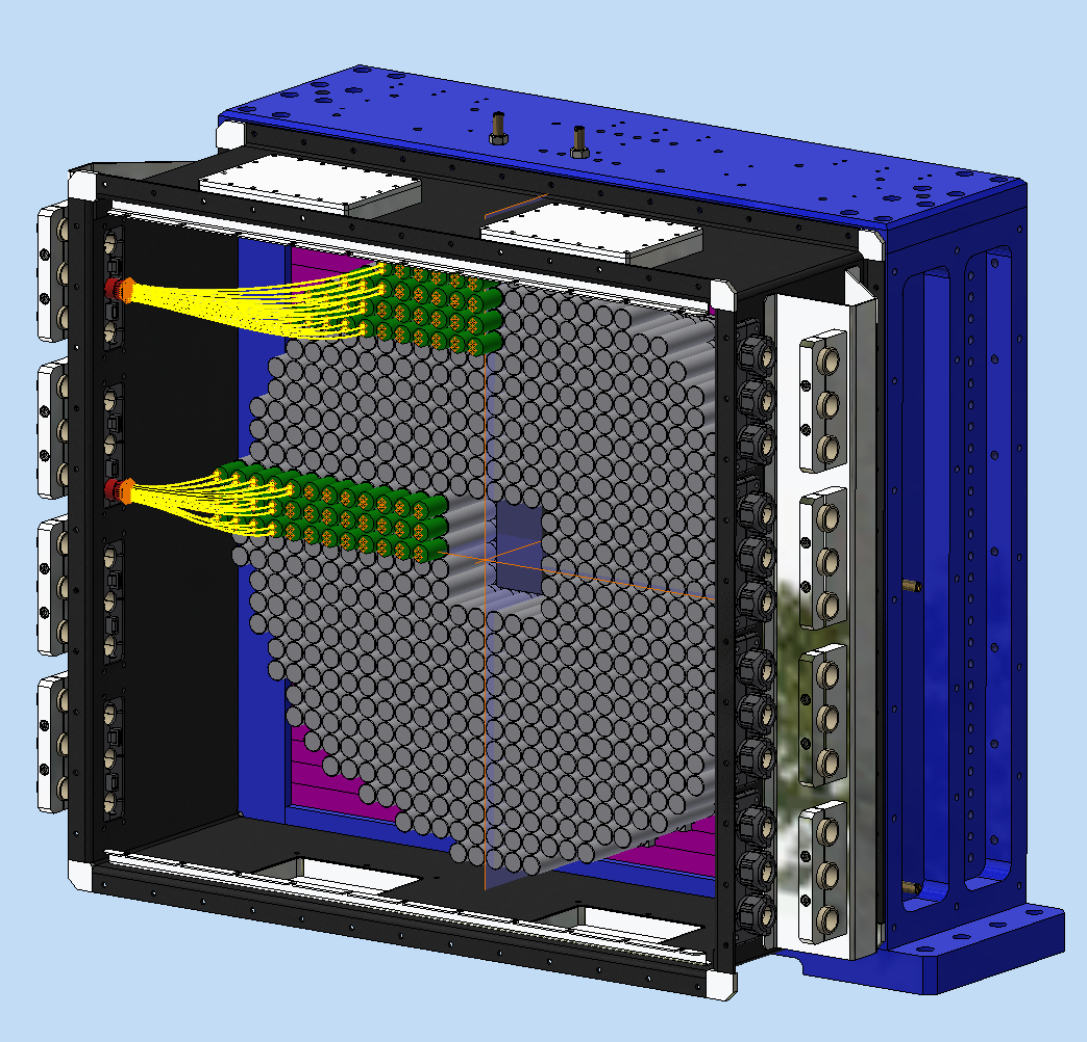}
\par\end{centering}
\caption{\label{fig:ECal} Rear view of the ECal without
back closing panel. The blue structure is the mechanical support, and the black case is the cover of the PMTs and their cables, which are 
shown only for few units. The purple components are plastic fillers (see text for more detail).}

\end{figure}

A CAD drawing of the calorimeter is presented in figure~\ref{fig:ECal}. The metallic support structure has a square shape chosen to ease the detector assembly.
The remaining free space between the frame and the crystals is filled
with plastic elements. This also puts BGO in contact with a low-density material instead of metal where photons could produce showers with higher probability. All HV and signal cables exit from
the back in groups of $64$, passing through two holders, the inner one light-tight.

Due to the low energy of the photons in PADME, any single crystal support structure could spoil the energy resolution by introducing dead materials. 
In assembling the calorimeter special care has been taken to control size differences among the SUs which were individually selected on the basis of their dimensions.  $50\,\mu\text{m}$ black Tedlar$^{\circledR}$~\cite{Tedlar} strips were used to 
compensate for differences in their heights.
To reduce light cross-talk, additional Tedlar$^{\circledR}$ strips were inserted vertically between adjacent crystals, while foils were places horizontally between consecutive layers.

\subsection{Scintillating units production\label{subsec:Scintillating-units-production}}

The process of building scintillating units for the ECal required different steps. First, the old reflective paint and photodiodes were removed
from the original L3 crystals. To recover performance deterioration
due to radiation damage, crystals underwent accelerated annealing at CERN's LAB 27:
\begin{enumerate}
\item crystals were heated from room temperature to $200\,\text{\textdegree C}$ over $3\,\text{h}$;
\item they were kept at $200\,\text{\textdegree C}$ for $6\,\text{h}$;
\item finally, they were cooled down from $200\,\text{\textdegree C}$ back to room temperature over about $1$ day.
\end{enumerate}

\begin{figure}
\begin{centering}
\includegraphics[width=0.49\textwidth]{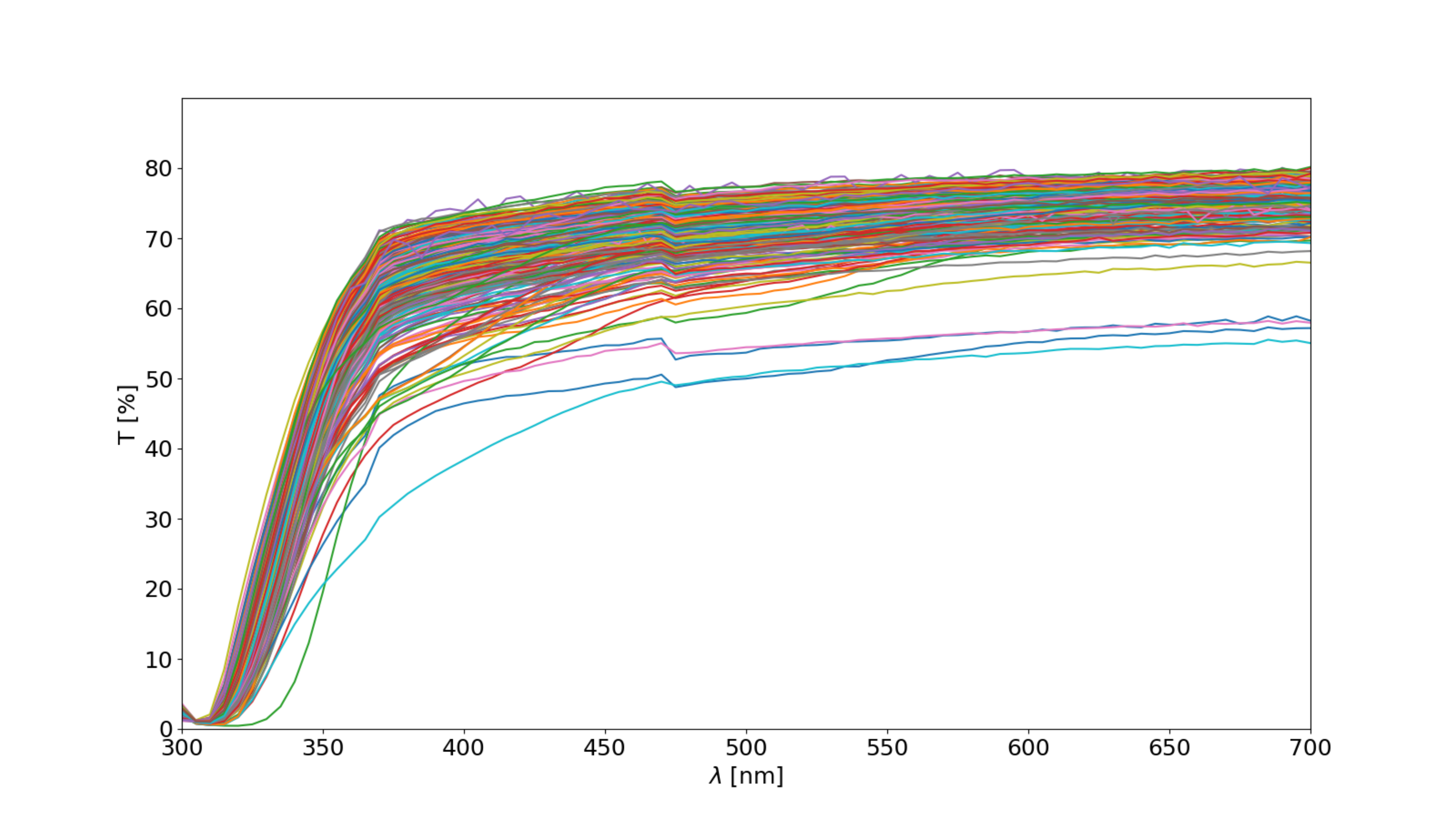}
\includegraphics[width=0.49\textwidth]{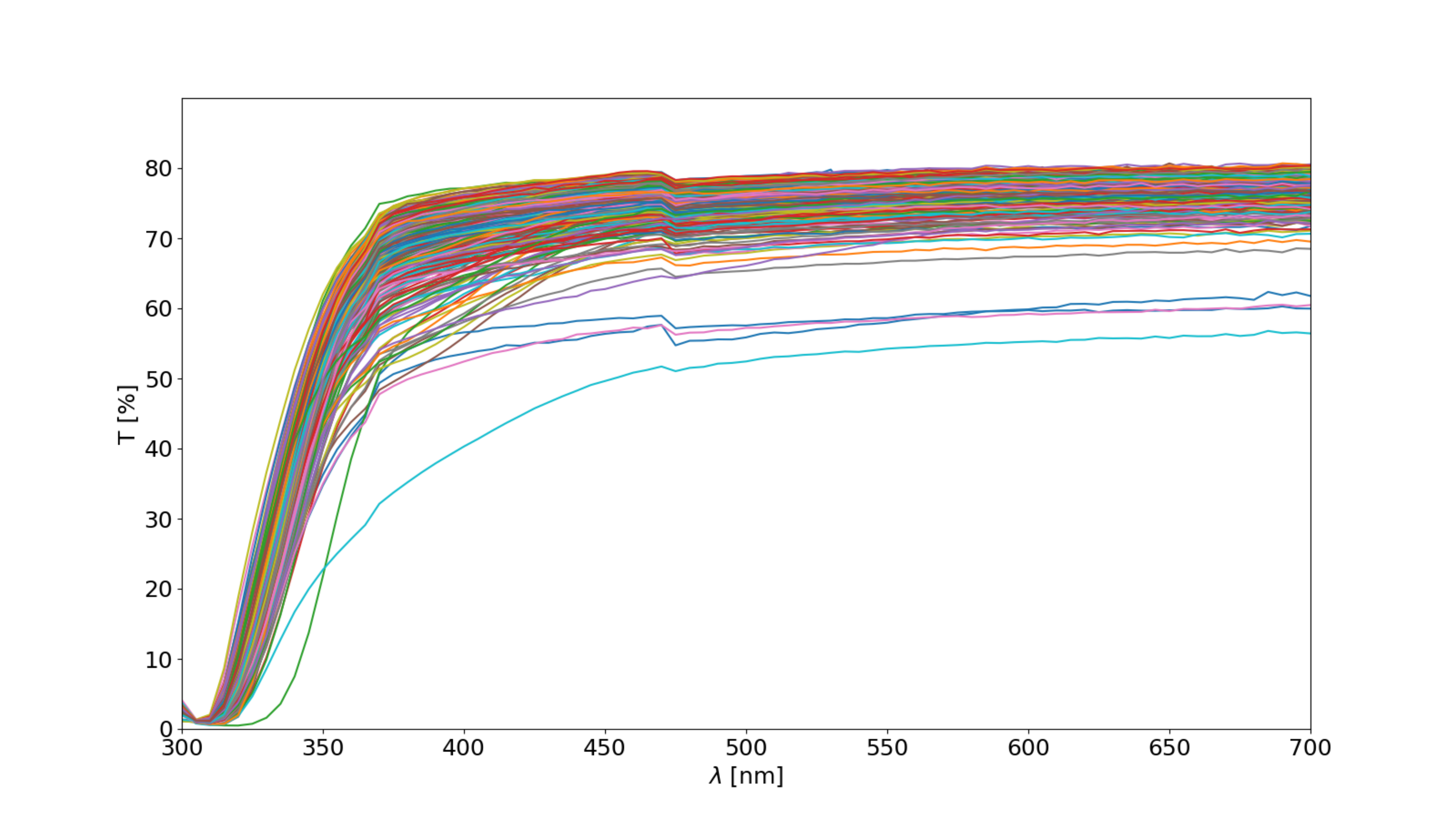}
\par\end{centering}
\caption{\label{fig:Transparency}
BGO transparency as a function of wavelength, before (left) and after (right) annealing. The improvement is visible especially at lower wavelengths.}
\end{figure}

Figure~\ref{fig:Transparency} shows measurements of the transparency before (left) and after (right) annealing. The improvement in transparency at lower wavelengths is clearly visible.

Since the original shape of the crystals was tapered pyramidal, they were cut at Gestione SILO (Italy)~\cite{SILO} to have rectangular faces. The same firm also glued the PMTs (using the ELJEN EJ-500 optical cement \cite{EJ-500}) and coated the BGO with three layers ($\approx60\,\mu\text{m}$) of ELJEN EJ-510 (a bright white diffusive paint with titanium dioxide pigments \cite{EJ-510}).

The PMTs used in the experiment are the HZC XP1911 type B \cite{HZC}, with a quantum efficiency of $21\%$ at $480\,\text{nm}$ (the BGO maximum emittance wavelength) and a $19\,\text{mm}$ diameter.
The gain curve of each PMT was studied before gluing by flashing a blue LED in front of the photocathode using different PMT HV values, and measuring the corresponding charge collected by each tube.

\subsection{Signal readout and trigger\label{subsec:Signal-readout}}

ECal PMT signals are digitised using CAEN V1742 boards~\cite{CAEN V1742,test V1742}. These host $4$ DRS4 ASICs (a switched-capacitor array sampling
chip) and provide a total of $32$ channels. Each channel has a dynamic
range of $1\,\text{V}$ with $12\text{-bit}$ precision. 

\begin{figure}
\begin{centering}
\includegraphics[width=0.6\textwidth]{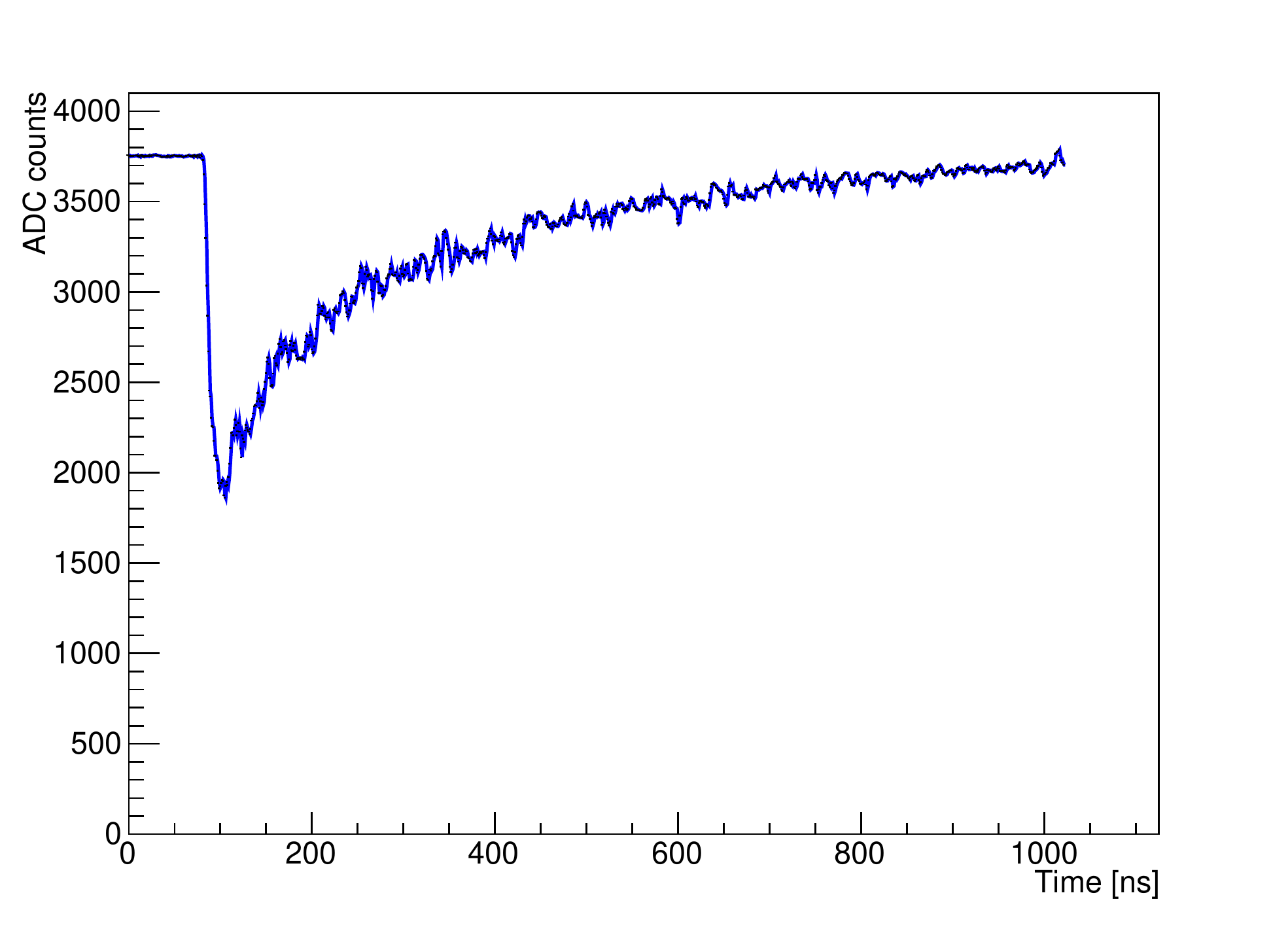}
\par\end{centering}
\caption{\label{fig:Pulse} A typical ECal signal digitized with a sampling frequency of $1\,\text{GS/s}$.}
\end{figure}

The board has selectable sampling frequency ($1\,\text{GS/s}$,
$2.5\,\text{GS/s}$, or $5\,\text{GS/s}$) and a memory depth of 1024 samples/channel. 
The selected sampling frequency for ECal signals is $1\,\text{GS/s}$,
to obtain a 1$\mu\text{s}$ long digitisation window able to match the long
decay time of BGO scintillation light. Figure~\ref{fig:Pulse} illustrates a typical calorimeter pulse.

For each digitisation window, some samples before the BGO signal leading
edge are collected to evaluate the pulse base-line, which is obtained as the mean of the first $50$ pre-pulse samples.
The total collected charge is then calculated summing over all the acquired samples and subtracting the base-line value.
Algorithms to correct for possible cuts on the pulse tail and for variation of the particle energy loss are also implemented.
\par
During the data-taking two different kind of triggers are used, calibration trigger and physics trigger:
\begin{enumerate}
\item  Cosmic ray (CR) trigger: provided by two scintillating slabs, placed above and below the ECal (see section~\ref{subsec:Cosmic-rays}). Its purpose is to check the SUs stability using signals from minimum ionizing particles (MIPs); 
\item Beam trigger: generated by the accelerator complex for each bunch, and used in data-taking for physics.
\end{enumerate}

\section{Scintillating units calibration and equalisation}
\label{calib}

SUs were calibrated with a $^{22}$Na source before mounting in the calorimeter. This was done to determine the charge vs HV curve used to set the SU voltage to the desired gain. 
After the installation of ECal in the experimental hall, the CR trigger was also used to check the response of the units to MIPs. Using CR events, it is possible to validate the $^{22}$Na calibration and to assess and improve the SUs 
equalisation.

\subsection{Pre-assembly calibration with $^{22}$Na}

\begin{figure}
\begin{centering}
\includegraphics[height=2cm]{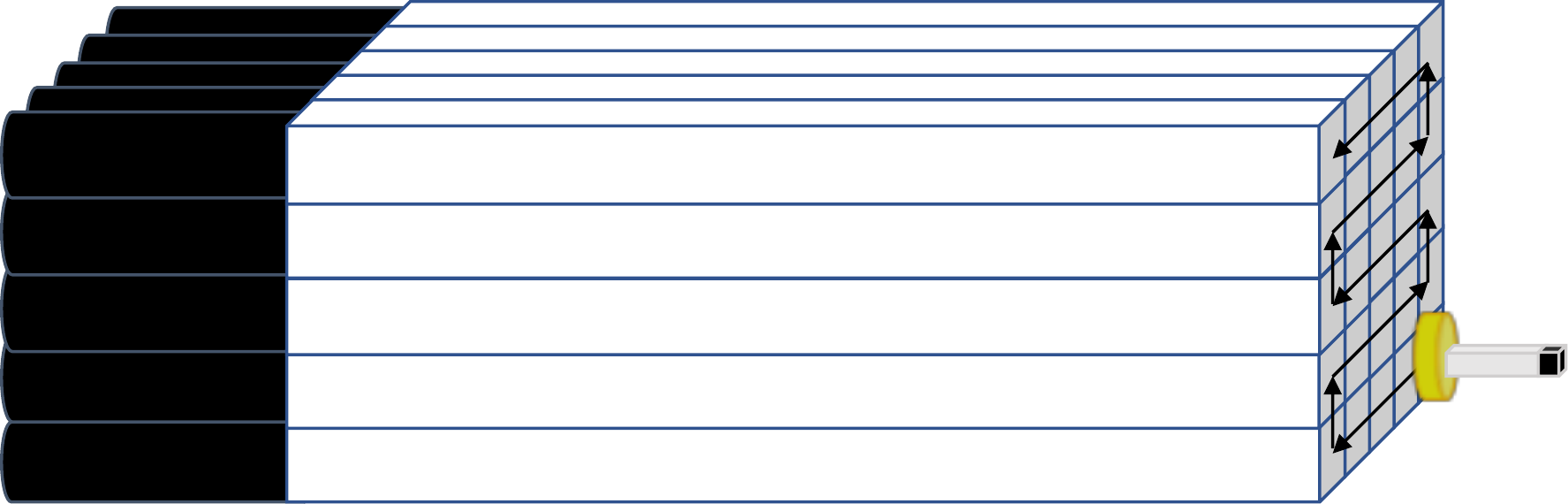}~~~\includegraphics[scale=0.27]{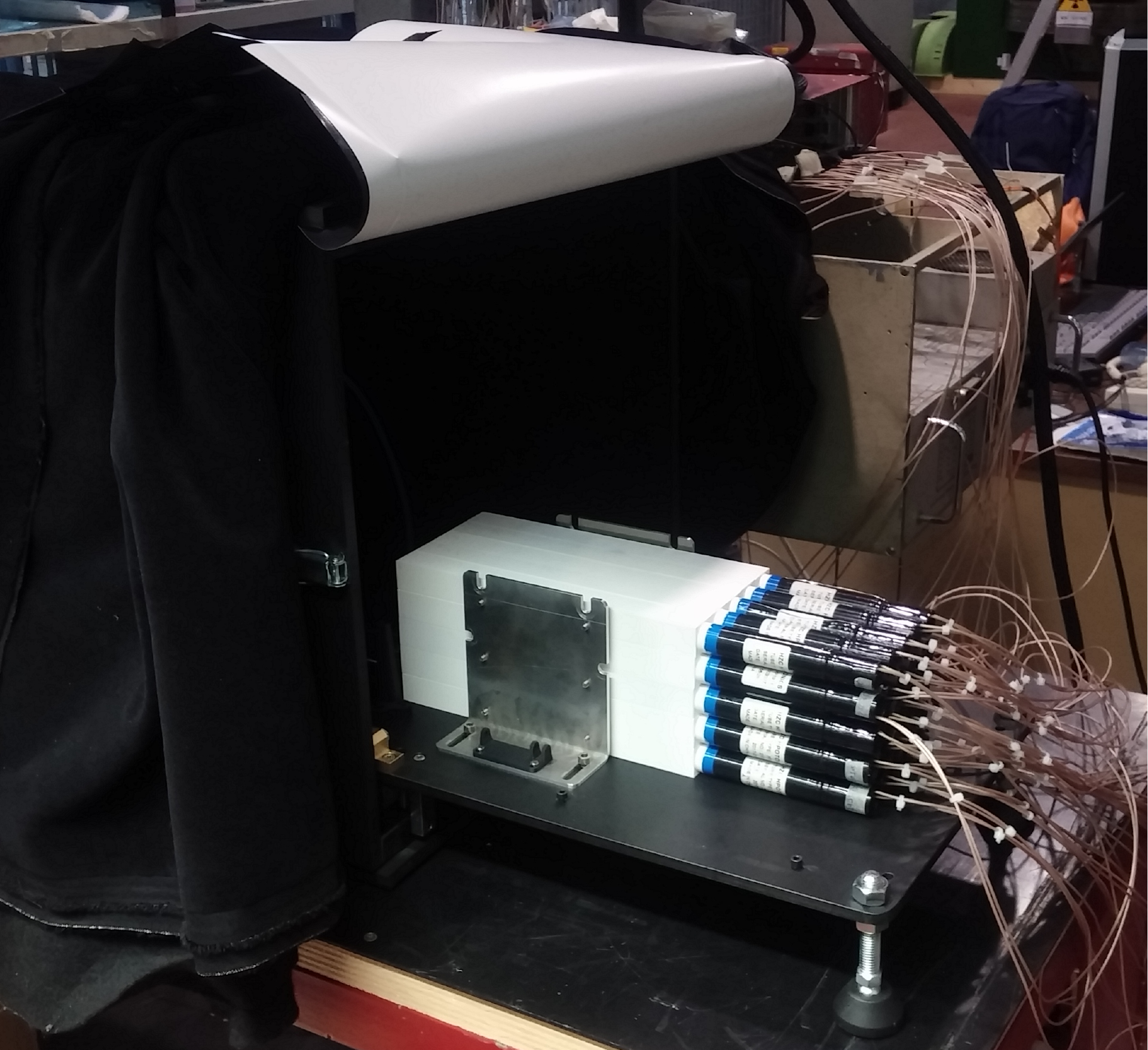}
\par\end{centering}
\caption{\label{fig:Na-22_setup}Setup for SUs calibration. Left: sketch of
the test stand (not to scale), with the $^{22}$Na source indicated
with a yellow disk (possible motions are also shown). A LYSO crystal
is placed on the opposite side of the source to provide the trigger
signal. Right: a photograph of the test stand, with a stack of SUs.}
\end{figure}

To select and equalise the response of each SU, a dedicated setup
was built exploiting the two back-to-back $511\,\text{keV}$ photons
emitted by a $^{22}$Na source. 
It allows the scan of a $5\times5$
SU matrix with the source moving in front of each crystal. Figure~\ref{fig:Na-22_setup}
(left) shows a drawing with the sodium path highlighted and figure~\ref{fig:Na-22_setup}
(right) shows a photograph of the test stand. A $3\times3\times20\,\text{mm}^{3}$
LYSO crystal, read out by a SiPM, is placed on the opposite side of the source with respect to the BGO and produces the trigger signal.
Data were collected at $10$ different voltages in the interval $[1100,1550]\,\text{V}$ in steps of $50\,\text{V}$, acquiring about $6000$ events per HV value.

\begin{figure}
\begin{centering}
\includegraphics[width=0.49\textwidth]{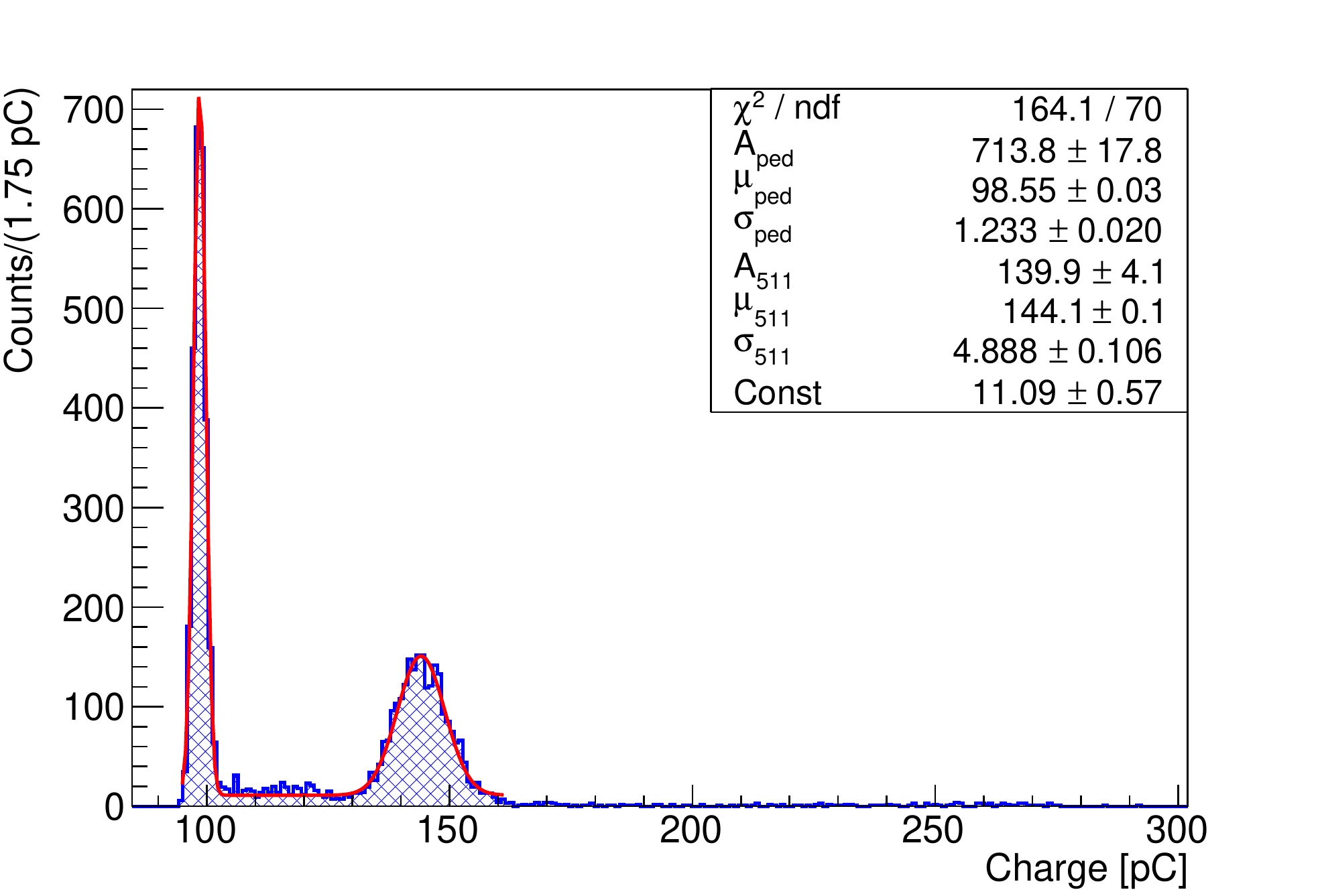}
\includegraphics[width=0.49\textwidth]{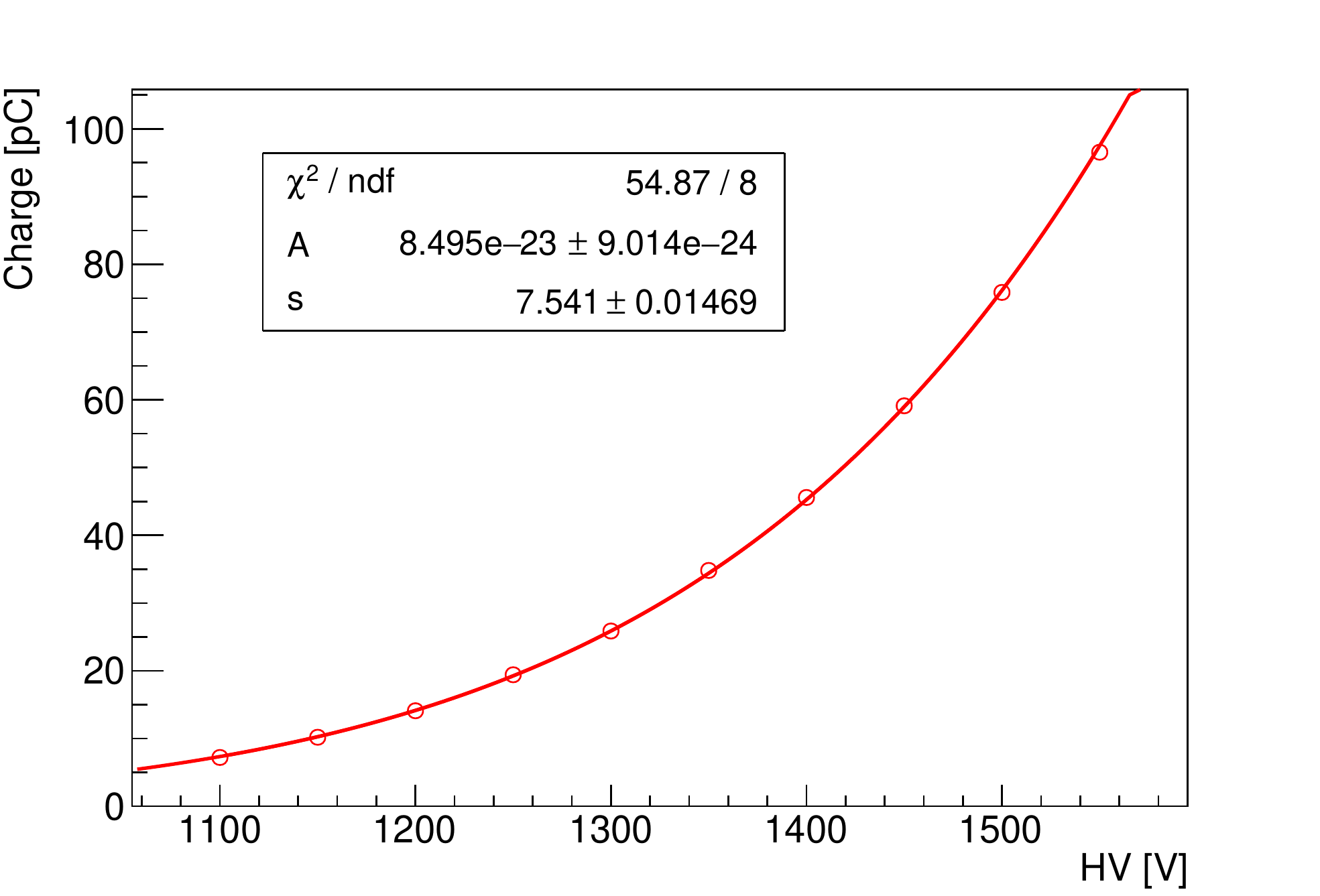}
\par\end{centering}
\caption{\label{fig:Charge_HV}Left: charge distribution of a SU at $1400\,\text{V}$
fit with a double Gaussian plus a flat function to reproduce the pedestal, the $511\,\text{keV}$ peak, and the constant background. 
Right: Measured charge as a function of HV with the gain curve fit $Q=A\cdot V^{s}$ superimposed.}
\end{figure}

An example of the charge distribution obtained for a unit at $1400\,\text{V}$
is presented in figure~\ref{fig:Charge_HV} (left). The pedestal and the
$511\,\text{keV}$ signal are clearly visible and both are fitted
with a Gaussian, while the approximately flat continuum is modelled with a constant function.

The collected charge, for each value of the HV, was determined as the difference between the $511\,\text{keV}$ peak and pedestal position. 
Figure~\ref{fig:Charge_HV} (right) presents the charge vs HV behaviour with the best-fit curve superimposed (same SU of figure~\ref{fig:Charge_HV} (left)). The fitting curve has the form
$Q=A\cdot V^{s}$, where $Q$ is charge, $V$ is voltage, and $A$ and
$s$ are free parameters.

\begin{figure}
\begin{centering}
\includegraphics[height=6cm]{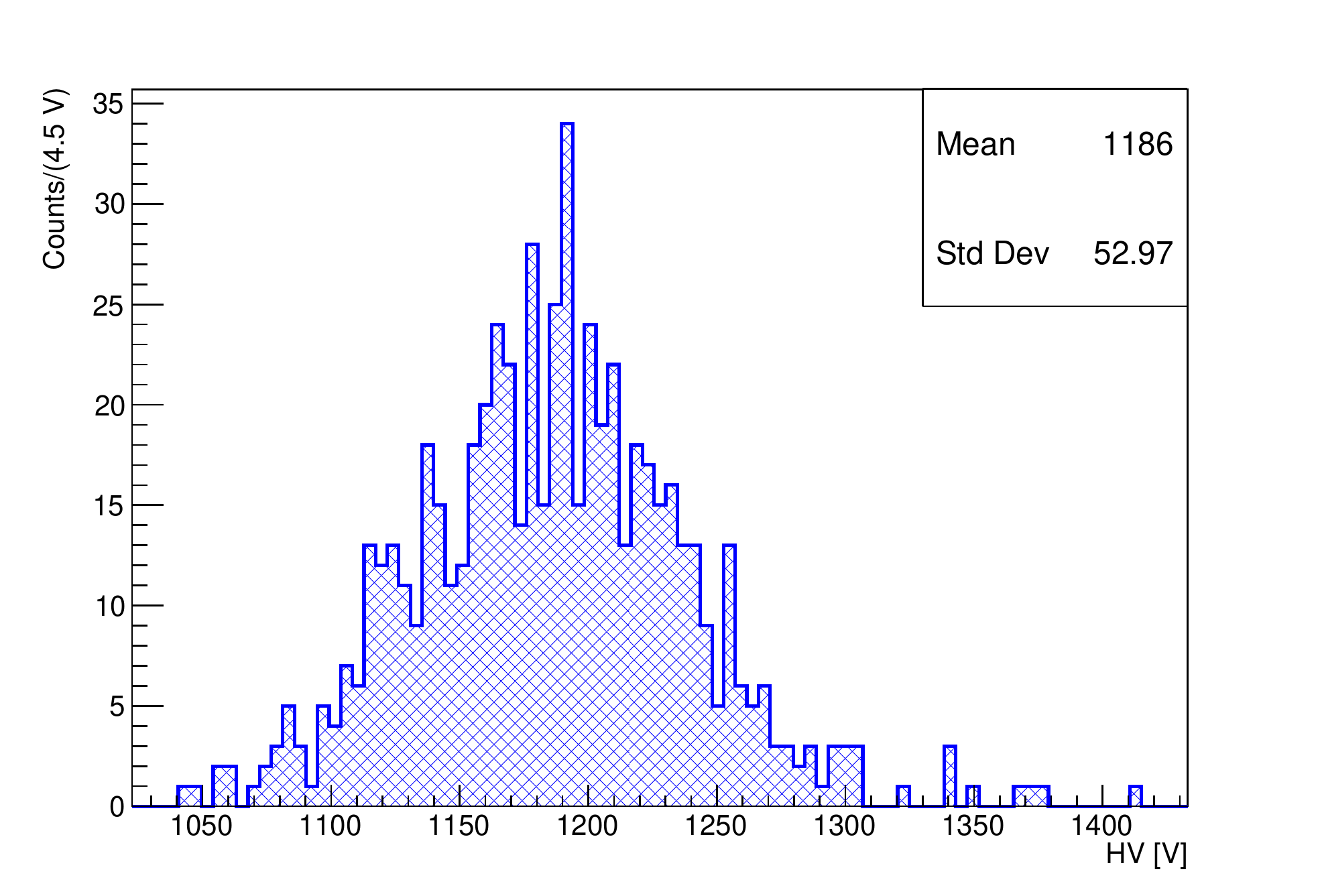}
\par\end{centering}
\caption{\label{fig:HV_distr}Distribution of PMT voltages used to set the gain of the corresponding SU to $15.3\,\text{pC/MeV}$.}
\end{figure}

In figure~\ref{fig:HV_distr}, the distribution of voltages needed to obtain a gain of $15.3\,\text{pC/MeV}$ is displayed for all $616$ PADME SUs. The distribution mean is $1186\,\text{V}$
with a standard deviation of $53\,\text{V}$. Even the largest value, $1411\,\text{V}$, is well below the safety operational maximum of $1700\,\text{V}$~\cite{HZC}. 

\begin{figure}
\begin{centering}
\includegraphics[height=6cm]{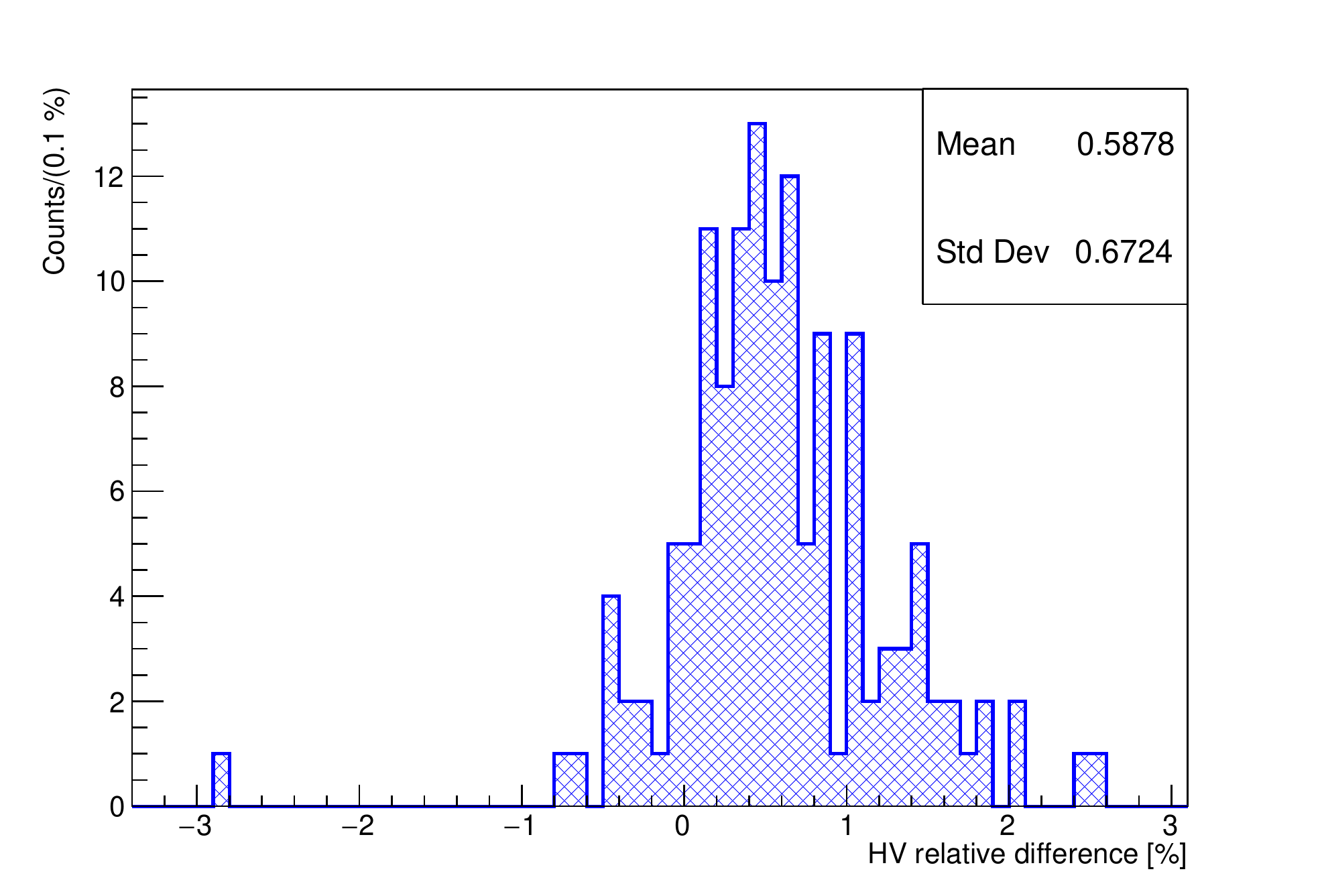}
\par\end{centering}
\caption{\label{fig:HV_rel_diff}Relative difference between voltage measurements for the $135$ SUs that were tested a second time to assess gain curve reproducibility. The required gain is $15.3\,\text{pC/MeV}$ in both cases.}
\end{figure}

To evaluate the reproducibility of calibrations, $135$
SUs ($22\%$ of the total) underwent a second, identical, measurement campaign. 
The relative difference between the two optimal HVs, given by $\tfrac{V_{1}-V_{2}}{\left(V_{1}+V_{2}\right)/2}$, remains below a few percent as shown in figure~\ref{fig:HV_rel_diff}. 
This difference also incorporates possible temperature variations between the two measurements. 

\subsection{Cosmic rays\label{subsec:Cosmic-rays}}

\begin{figure}
\begin{centering}
\includegraphics[height=6cm]{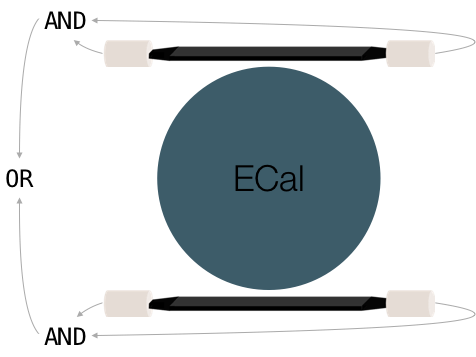}
\par\end{centering}
\caption{\label{fig:CRs_trigger}Structure and logic of the ECal CR trigger.
It consists of two plastic scintillator slabs, one above and one below
the ECal. Each bar is read out by two PMTs, one per side, set
in logic AND. The logic OR of the two ANDs gives the trigger signal.}
\end{figure}

Since the start of data-taking, the calorimeter has been operated
with voltages corresponding to a gain of $15.3\,\text{pC/MeV}$. As
shown in figure~\ref{fig:CRs_trigger}, the calorimeter is equipped with
a CR trigger, made by two plastic scintillator slabs, one above and
one below the ECal, and each read out by two PMTs (one per side). The trigger logic implemented is also detailed in the figure.

\begin{figure}
\begin{centering}
\includegraphics[width=0.65\textwidth]{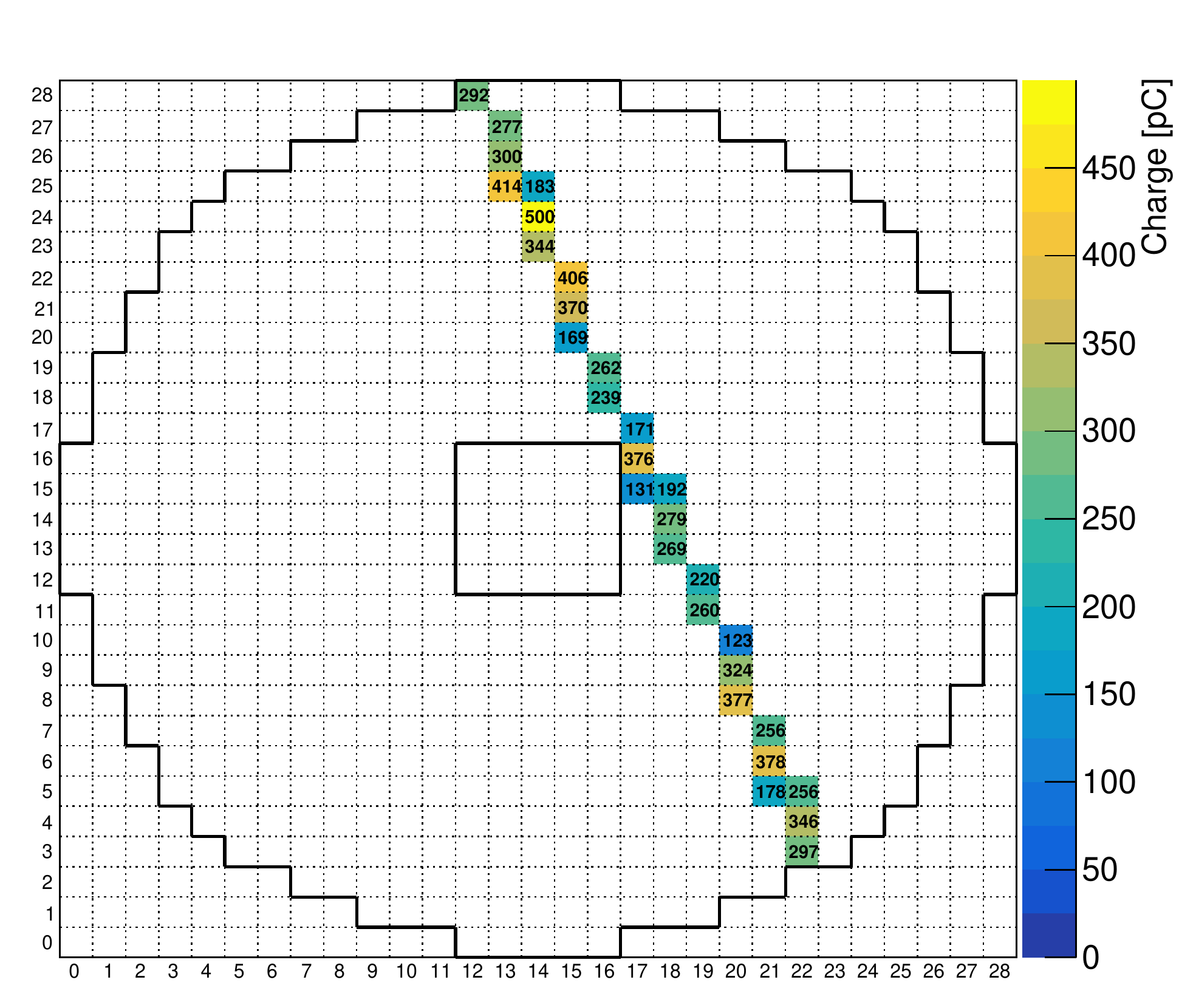}
\par\end{centering}
\caption{\label{fig:CR_ECal}A cosmic ray passing through the ECal. The color
scale and numbers inside the squares represent the charge collected
by the SU.}
\end{figure}

Figure~\ref{fig:CR_ECal} gives an example of a cosmic ray passing through the
entire calorimeter. The numbers in the various positions indicate the charge collected by each SU, in pC.

The calorimeter equalisation is checked by studying the charge distributions obtained from cosmic rays in the various SUs. For a more precise evaluation, only cosmics crossing crystals vertically are considered. This is defined by three conditions:
\begin{itemize}
\item a cosmic ray passes through three SUs aligned in a column;
\item there are no other signals in the three rows of the SUs under consideration;
\item only the pulse from the central SU is considered for the calculation.
\end{itemize}

This ensures that the cosmic ray releases little energy in adjacent crystals. For example, in figure~\ref{fig:CR_ECal}, only SUs in position $(15,21)$ and $(20,9)$ satisfy these requirements. For peripheral units without any SUs directly above or below ($68$, corresponding to $11\%$ of the total), the two crystals below or above are used, respectively.

\begin{figure}
\begin{centering}
\includegraphics[height=6cm]{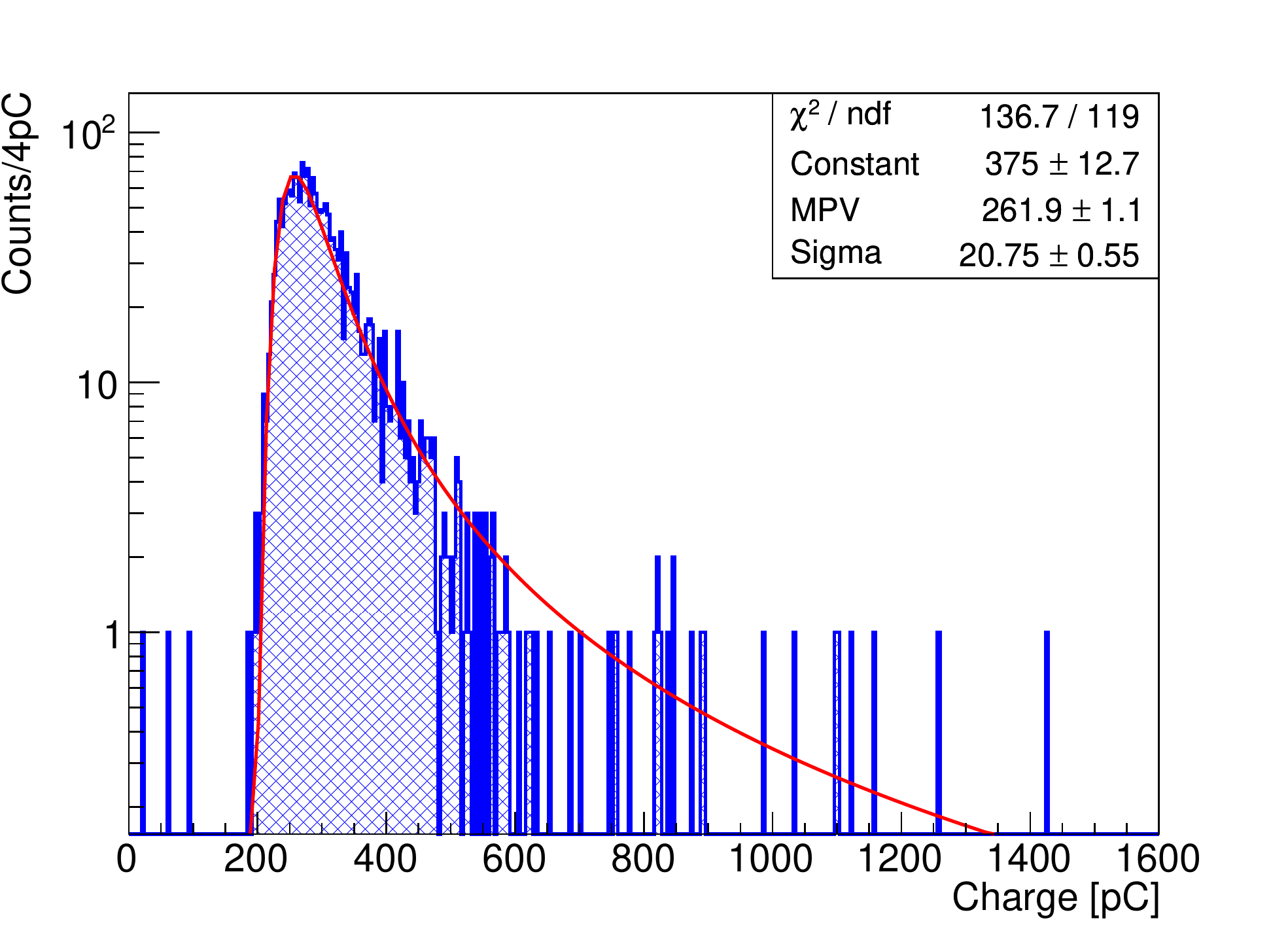}
\par\end{centering}
\caption{\label{fig:CR_charge_distr}Example of charge distribution given by CRs passing vertically through a SU. A Landau fit is superimposed.}
\end{figure}

An example of the charge distribution produced by this selection, with a superimposed Landau fit, is given 
in figure~\ref{fig:CR_charge_distr}. Landau parameters are those defined by the ROOT package \cite{root} and the algorithm is from CERNLIB G110 DENLAN, which is based on~\cite{Landau}.

\begin{figure}
\begin{centering}
\includegraphics[height=6cm]{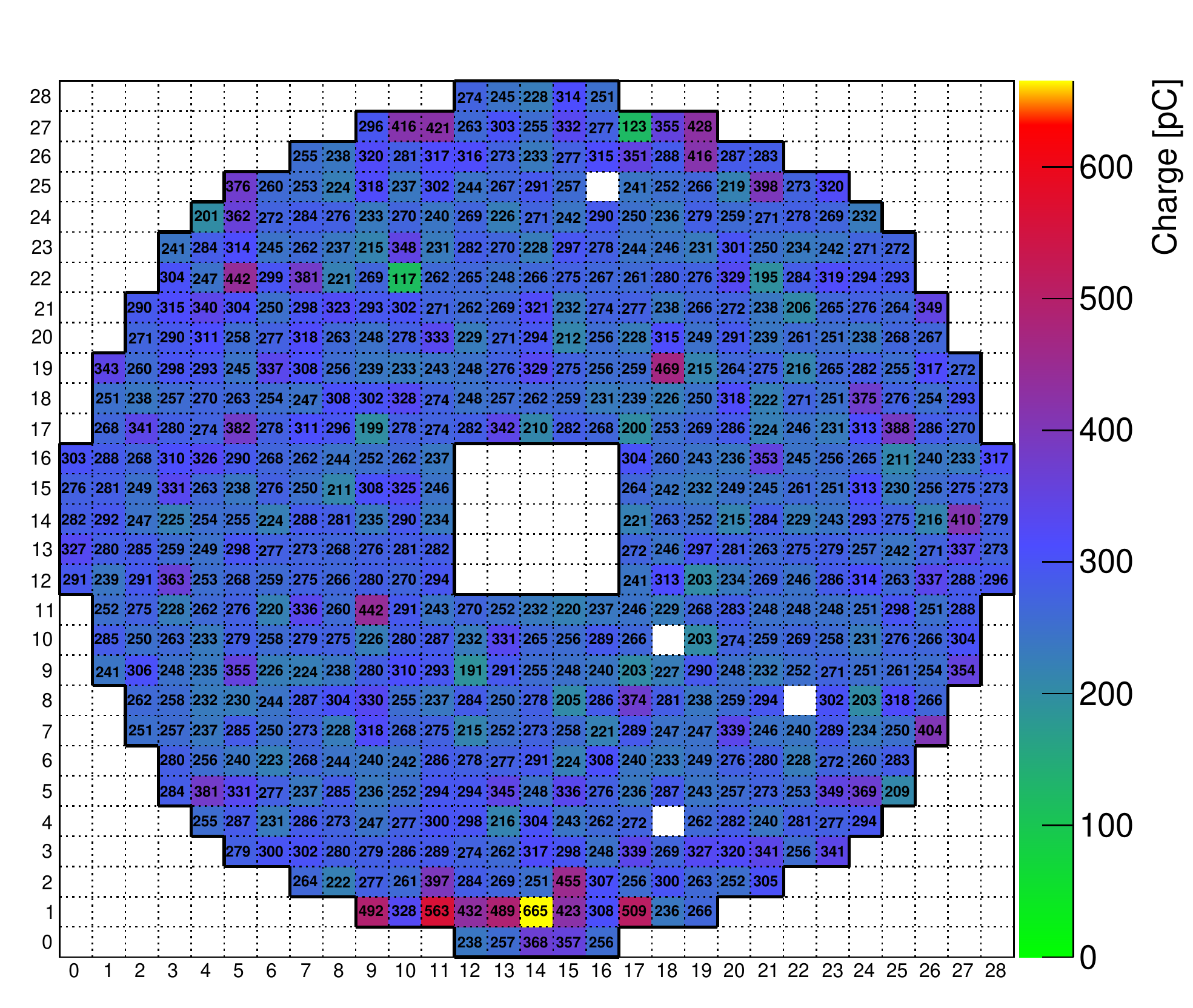}\includegraphics[height=4.4cm]{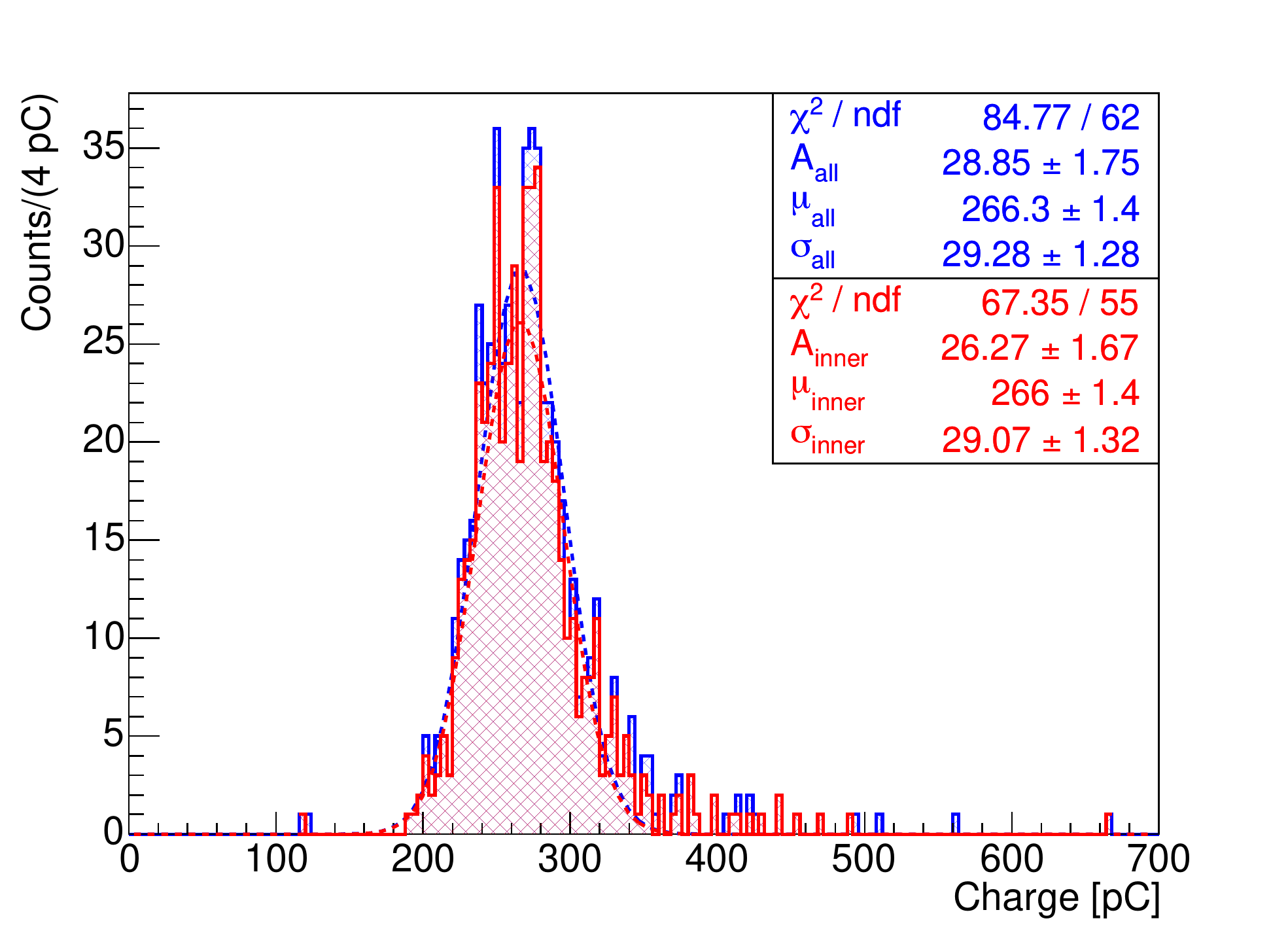}
\par\end{centering}
\caption{\label{fig:CR_MPV}MPVs obtained from Landau fits
to the charge distributions produced by CRs passing vertically through SUs. Left:
map of charge measurement in the ECal. The $4$ white squares correspond to non-operational SUs. Right: MPV distributions considering
(blue) and not considering (red) the $68$ edge crystals.} 
\end{figure}

Figure~\ref{fig:CR_MPV} presents the most probable values (MPVs) from Landau fits performed on charge distributions obtained from CR data collected over three days. On the left, the 2D distribution (with values and color scale) is shown and on the right two histograms are superimposed: the blue histogram corresponds to all active SUs while the
red one excludes the $68$ peripheral units. The means of the two Gaussian fits are compatible: $(266.3\pm1.4)\,\text{pC}$ and $(266.0\pm1.4)\,\text{pC}$, respectively. 
Taking into account the Gaussian for all the units, the ECal equalisation achieved using the gain curves from the $^{22}$Na source indicates a signal spread of $(10.99\pm0.48)\%$, where the spread is expressed as the ratio of the sigma and of the mean of the Gaussian fit.

Empty squares on the left plot of figure~\ref{fig:CR_MPV} indicate $4$ non-operational channels. 
If one of these belongs to the triplet needed to define the verticality condition, it is skipped and the one above or below is used.
This study also helps to improve the calorimeter energy resolution. Assuming that, on average, CRs release the same amount of energy in all crystals, the MPV of each charge distribution can be used as a normalisation term, by simply dividing each pulse integral by the MPV of the corresponding ECal channel.

\begin{figure}
\begin{centering}
\includegraphics[height=6cm]{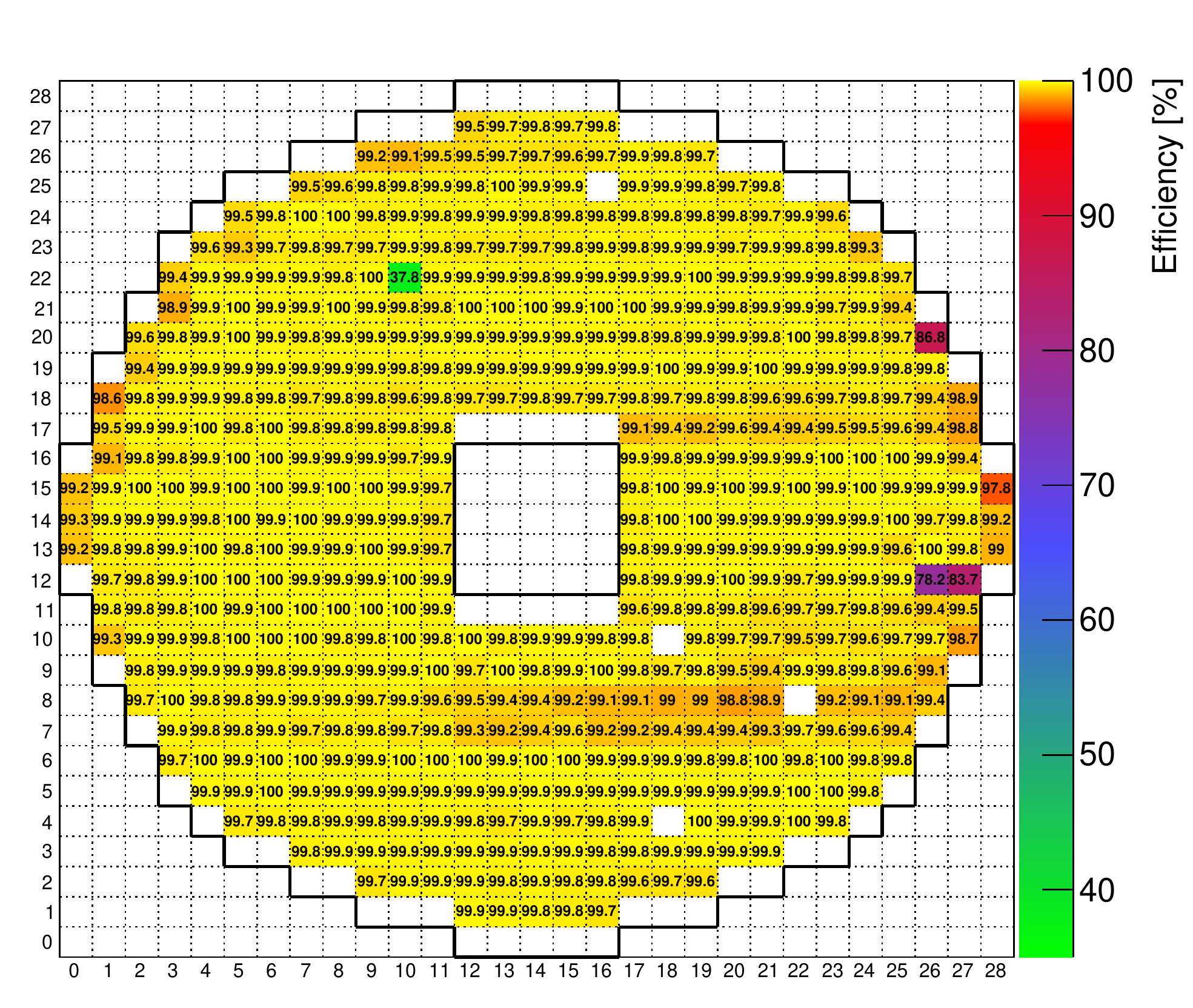}\includegraphics[height=4.5cm]{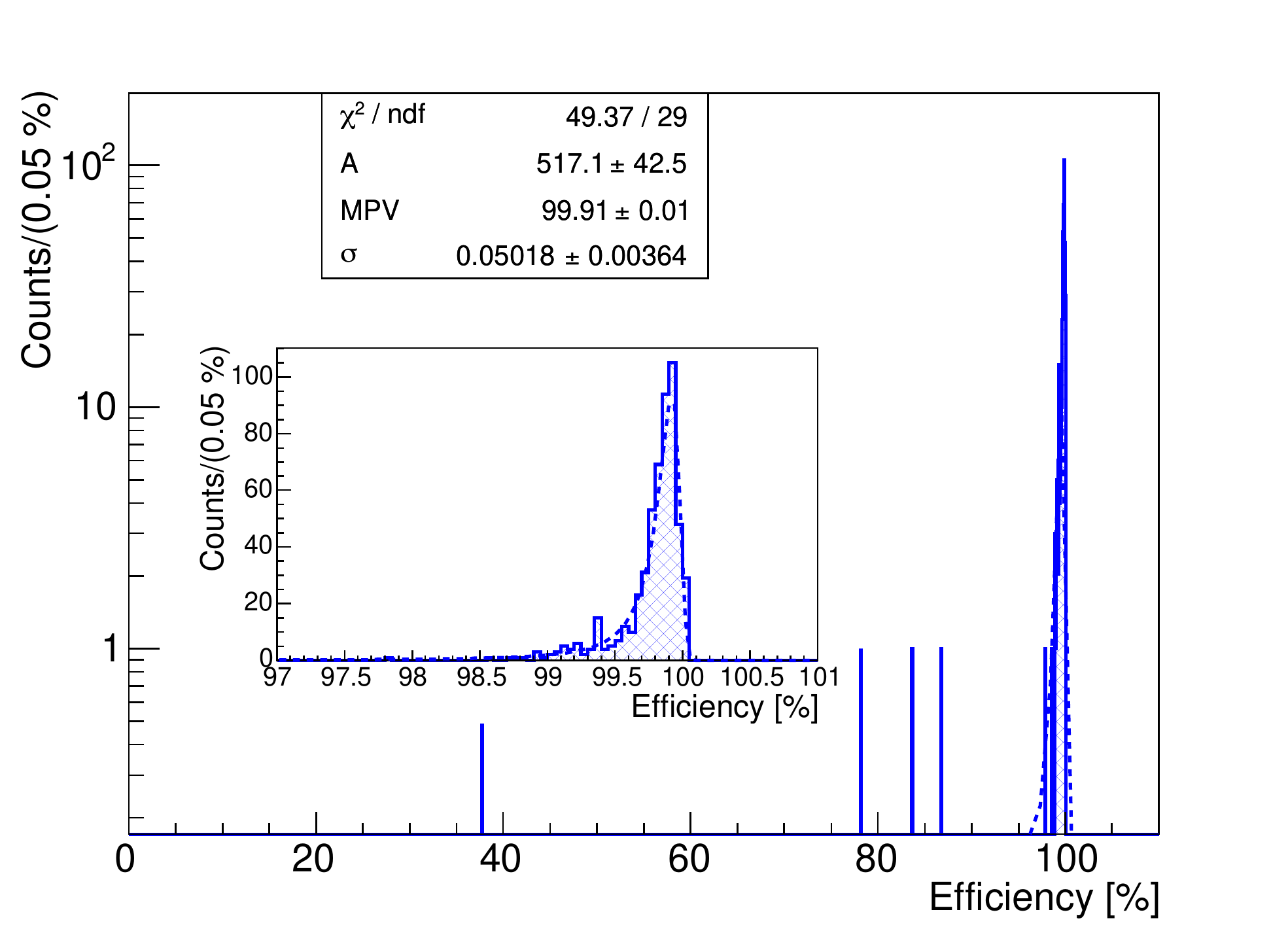}
\par\end{centering}
\caption{\label{fig:Efficiency}Efficiency of SUs, evaluated using CR data. Left: ECal efficiency map.
Right: efficiency distribution with a reversed Landau fit superimposed (see text). In the inset, a zoom on the region $[97,101]\%$. Both plots exclude the $4$ non-operational units and the $68$ edge units.}
\end{figure}

Finally, CRs are also used to evaluate the SU efficiency. 
With the same triplets defined in the previous analysis, the efficiency of the central SU can be determined.
The efficiency is defined as the ratio between the number of events with hits in all three cells and that with hits only in the two external ones. 
Figure~\ref{fig:Efficiency} (left) shows the 2D distribution of ECal efficiencies.
They have been evaluated using a CR data set taken over three days.  Figure~\ref{fig:Efficiency} (right) reports the measured efficiencies with a zoom on the interval $[97,101]\%$ in the inset. A reverse Landau fit is superimposed on both distributions.

\begin{figure}
\begin{centering}
\includegraphics[height=5cm]{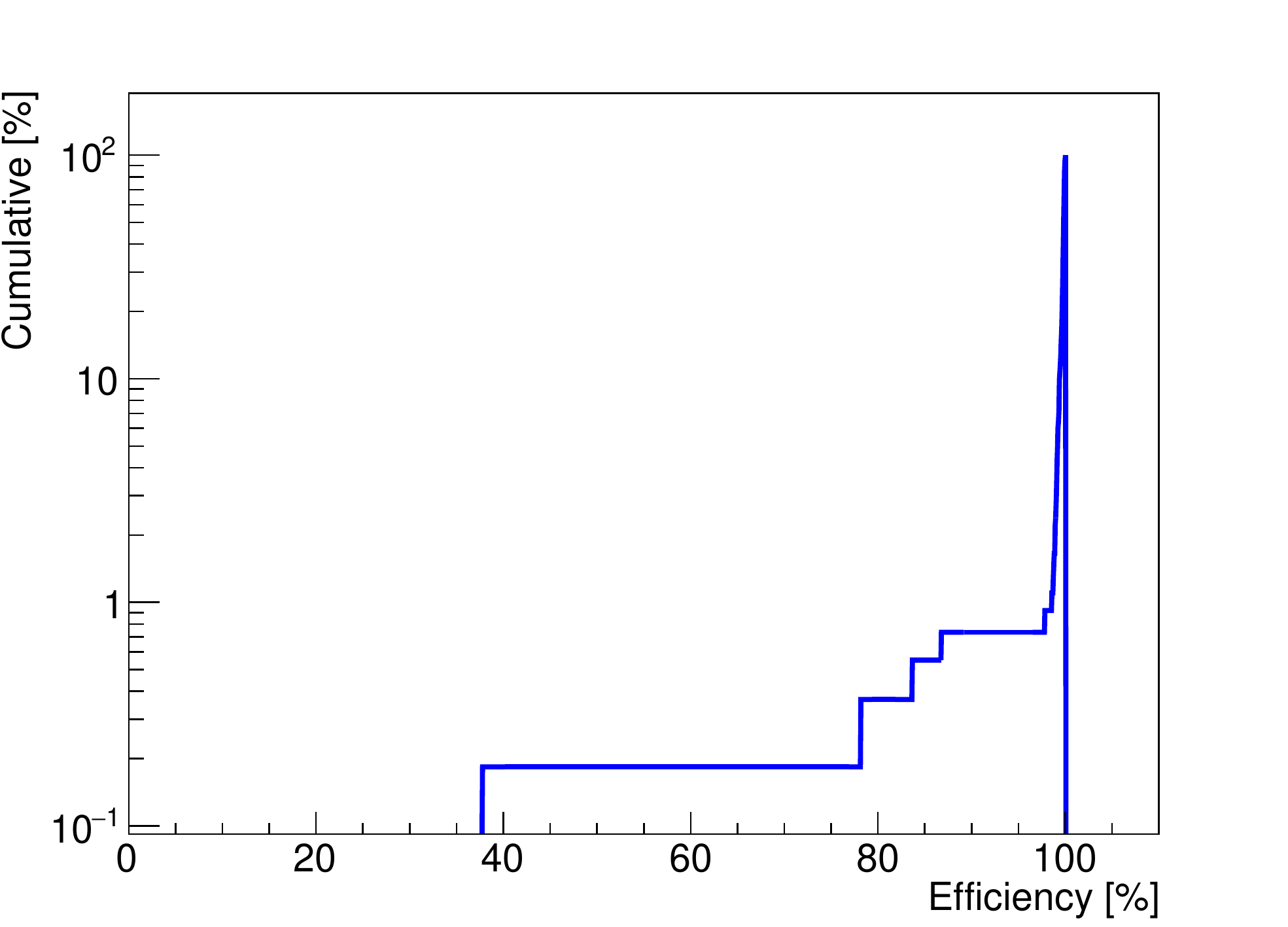}
\par\end{centering}
\caption{\label{fig:EfficiencyCumulative}Cumulative efficiency distribution of figure~\ref{fig:Efficiency}, in percent. The percentage is given without considering the 4 non-operational units and the 68 edge units.}
\end{figure}

Figure~\ref{fig:EfficiencyCumulative} reports the cumulative efficiency distribution of SUs. The percentage is given without considering the 4 non-operational units and the $68$ edge SUs.

\subsection{Calorimeter performance with positron beam}

Before ECal construction, several tests were performed on a prototype consisting in a $5\times5$ matrix of BGO crystals ($2.0\times2.0\times22.0\,\text{cm}^{3}$) read out by XP1912 HZC PMTs. 
In particular, single-positron beams of different energies were fired at the central
crystal of the matrix. These measurements reveal an energy resolution compatible with the desired performance~\cite{ECal prototype}:

$$\frac{\sigma(E)}{E}=\frac{2.0\%}{\sqrt{E[\text{GeV}]}}\oplus\frac{0.003\%}{E[\text{GeV}]}\oplus1.1\%.$$

\begin{figure}
\begin{centering}
\includegraphics[height=6cm]{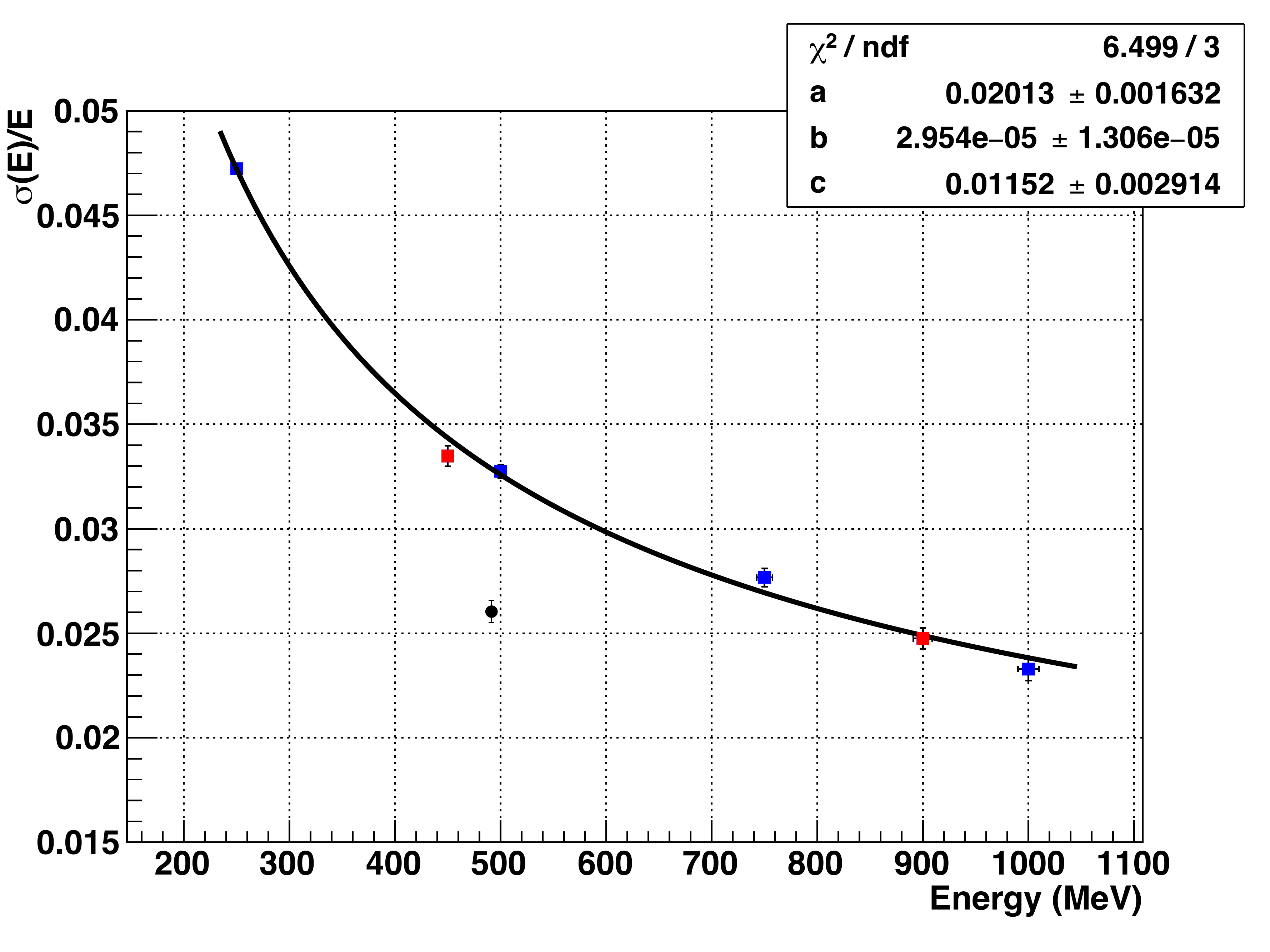}
\par\end{centering}
\caption{\label{fig:EnergyReso} Energy resolution measured with the ECal prototype (red and blue points) \cite{ECal prototype} compared with the ECal result on special run data set (black point).} 
\end{figure}

Figure~\ref{fig:EnergyReso} presents the energy resolution measured during these tests. Blue (red) squares identify results obtained with a $250\,\text{MeV}$ ($450\,\text{MeV}$) beam energy. 
During the same test the charge response as a function of the deposited energy has been shown to be linear in within $2\%$ up to $1\,\text{GeV}$.

To evaluate the energy resolution of ECal  a special run was performed with
a $490\,\text{MeV}$ positron beam energy (multiplicity ${\approx}1$) fired directly at the calorimeter. A $5\times5$ cluster, around the crystal with the largest energy deposit, was considered to simulate the prototype condition (other clusterization algorithms implemented by the PADME collaboration are discussed in~\cite{Clusters}). The measured energy resolution turned out to be $2.62\pm0.05\,(\text{stat})\%$, as shown by the black circle in figure~\ref{fig:EnergyReso}. 
This improvement was partially expected since the prototype detector was slightly different from the final calorimeter.
Several factors contributed to improve the performance of the ECal with respect to the prototype:
\begin{itemize}
\item The ECal has longer crystals and therefore a better energy containment;
\item PMTs are glued to the BGO while in the prototype they were optically coupled with optical grease;
\item ECal crystals are painted and not simply wrapped with PTFE tape;
\item The ECal SUs responses are equalised a priori while at the test beam all SUs were operated at the same voltage of $1100\,\text{V}$.
\end{itemize}

The beam characteristics also contributes to the measured energy resolution: the beam used for the prototype test was not optimised as in the standard data-taking. In fact, in order to be able to change its energy, electrons produced on a secondary target were used. This was necessary since the main user of the LINAC primary beam was the DA$\Phi$NE LNF collider requiring a fixed energy (550 MeV). A primary beam, $10^{9}$ particles/bunch, was hitting a secondary target and electrons  of desired energy, produced by the electromagnetic showers, were then selected with a dipole magnet. 
Due to the high multiplicity, the showers also produced many photons that could reach the prototype under test, inducing background.
Conversely, in the ECal test, the beam was a primary beam consisting only of $490\,\text{MeV}$ positrons. In this case the LINAC gun was off and only few electrons were accelerated, the ones produced by the gun's dark current in phase with the accelerating field. This very low multiplicity helped reducing the beam-induced photon background reaching the ECal improving the measured energy resolution.

\section{Conclusions}
The primary goal of the PADME experiment is to measure with high efficiency and precision the SM photons produced by the reaction
 $e^{+}+e^{-}\rightarrow \gamma + A'$, where the dark photon  $A'$ signal appears as missing mass. This is obtained by means of an electromagnetic calorimeter made of 616 BGO crystals.
 
Each SU was calibrated with a $^{22}$Na source before installation in order to set the working point to $15.3\,\text{pC/MeV}$. Owing to the implementation of a cosmic ray trigger, the calibration of SUs was continuously monitored during physics data-taking. The same system also allowed SUs equalisation and efficiency evaluation. The measured gain spread is $(10.99\pm0.48)\%$, while the efficiency is $\geq98\%$ for $99.1\%$ of the channels.

Finally, to check the energy resolution, a dedicated measurement was performed by firing a single positron beam of $490\,\text{MeV}$ energy at the calorimeter. The obtained resolution is $2.62\pm0.05\,(\text{stat})\%$, with an evident 
improvement compared to the values obtained with a calorimeter prototype.

\acknowledgments
The PADME collaboration wants to warmly thank the BTF and LINAC teams, for the excellent quality of the beam provided during the test run.
The PADME collaboration wants to thanks the technical staff of LNF for the work done. Among the others a special mention goes to R. Lenci, and G. Papalino for their help in assembling and cabling the detector.

University of Sofia group is partially supported under BG-NSF DN08-14/14.12.2016 and the MoU SU -- LNF-INFN 70-06-497/07-10-2014.


\begin{thebibliography}{99}


\bibitem{DP 1}P. Galison and A. Manohar, \emph{Two Z's or not two Z's?}, 
\emph{Phys. Lett. B} {\bf136} (1984) 279

\bibitem{DP 2}B. Holdom, \emph{Two U(1)'s and $\epsilon$ charge shifts}, 
\emph{Phys. Lett. B} {\bf 166} (1986) 196

\bibitem{BTF1}G. Mazzitelli et al., \emph{Commissioning 
of the DA$\mathit{\Phi}$NE beam test facility}, 
\emph{Nucl. Instrum. Methods A} {\bf 515} (2003) 524

\bibitem{BTF2}P. Valente et al., \emph{Linear Accelerator Test Facility
at LNF Conceptual Design Report},
\emph{INFN-16-04-LNF}, arXiv:1603.05651 (2016)

\bibitem{Target}R. Assiro et al., \emph{Performance of 
the diamond active target prototype for the PADME 
experiment at the DA$\mathit{\Phi}$NE BTF},
\emph{Nucl. Instrum. Methods A} {\bf 898} (2018) 105

\bibitem{Veto1}F. Ferrarotto et al., \emph{Performance 
of the Prototype of the Charged-Particle Veto System of
the PADME Experiment}, 
\emph{IEEE Trans. Nucl. Sci.} {\bf 65} (2018) 2029

\bibitem{Veto2}S. Ivanov and V. Kozhuharov, \emph{The 
charged particle veto system of the PADME experiment}, 
\emph{AIP Conf. Proc.} {\bf 2075} (2019) 080005

\bibitem{SAC}A. Frankenthal et al., \emph{Characterization 
and performance of PADME's Cherenkov-based
small-angle calorimeter},
\emph{Nucl. Instrum. and Methods A} {\bf 919} (2019) 89

\bibitem{PADME}M. Raggi and V. Kozhuharov, \emph{Proposal 
to Search for a Dark Photon in Positron on Target Collisions 
at DA$\mathit{\Phi}$NE Linac}, 
\emph{Adv. High Energy Phys.} {\bf 2014} (2014) 959802

\bibitem{L3}B. Adeva et al., \emph{The construction of the L3 experiment}, 
\emph{Nucl. Instrum. Methods A} {\bf 289} (1990) 35

\bibitem{PDG} P.A. Zyla et al. (Particle Data Group), to be published in Prog. Theor. Exp. Phys. 2020, 083C01 (2020).

\bibitem{Tedlar}\url{https://www.dupont.com/content/dam/dupont/amer/us/en/photovoltaic/public/documents/DEC\_Tedlar\_GeneralProperties.pdf}

\bibitem{SILO}\url{https://www.gestionesilo.it/}

\bibitem{EJ-500}\url{http://www.ggg-tech.co.jp/maker/eljen/ej-500.html}

\bibitem{EJ-510}\url{http://www.ggg-tech.co.jp/maker/eljen/ej-510.html}

\bibitem{HZC}\url{http://www.hzcphotonics.com/en\_introduction\%20of\%20products.html}

\bibitem{CAEN V1742}\url{https://www.caen.it/products/v1742/}

\bibitem{test V1742}E. Leonardi et al., \emph{Development 
and test of a DRS4-based DAQ system for the PADME 
experiment at the DA$\mathit{\Phi}$NE BTF},
\emph{J. Phys. Conf. Ser.} {\bf 898} (2017) 032024

\bibitem{root} \emph{ROOT Data Analysis Framework}, \url{https://root.cern.ch//}

\bibitem{Landau}K.S. K\"{o}lbig and B.Schorr, \emph{A program package for the Landau distribution},
\emph{ Computer Phys. Comm.} {\bf 31} (1984) 97 [Erratum-ibid. 178 (2008) 972]

\bibitem{ECal prototype}M. Raggi et al., \emph{Performance of the
PADME Calorimeter prototype at the DA$\mathit{\Phi}$NE BTF},
\emph{Nucl. Instrum. Methods A} {\bf 862} (2017) 31

\bibitem{Clusters}E. Leonardi, G. Piperno and M. Raggi, \emph{Evaluation 
of clustering algorithms at the $< 1\,GeV$ energy
scale for the electromagnetic calorimeter of the PADME
experiment},
\emph{J. Phys. Conf. Ser.} {\bf 898} (2017) 072019


\end{thebibliography}
\end{document}